\documentclass[rotate,useAMS,usenatbib]{mn2e}

\usepackage{graphicx}
\usepackage{graphics}
\usepackage{epsfig}
\usepackage{rotate}

\title[Boxy/peanut ``bulges'']  
{Boxy/peanut ``bulges'': comparing the structure of galaxies with the 
underlying families of periodic orbits.}
 
\author[Patsis \& Xilouris]
  {P. A. Patsis$^{1}$\thanks{E-mail:ppatsis@cc.uoa.gr(PAP)},
  E. M. Xilouris$^{2}$\thanks{E-mail:xilouris@astro.noa.gr(EMX)}\\   
  $^1$ Research Center for Astronomy, Academy of Athens, Soranou Efessiou 4,
  GR-11527, Athens, Greece\\ 
  $^2$ Institute of Astronomy \& Astrophysics, National Observatory of
  Athens, I. Metaxa \& V. Pavlou, GR-15236, Athens, Greece}
\begin{document}
\maketitle  
 
\begin{abstract}
  The vertical profiles of disc galaxies are built by the material trapped
  around stable periodic orbits, which form their ``skeletons''. According to
  this, the knowledge of the stability of the main families of periodic orbits
  in appropriate 3D models, can predict possible morphologies for edge-on disc
  galaxies. In a pilot survey we compare the orbital structures which lead to
  the appearance of ``peanuts'' and ``X''-like features with the edge-on
  profiles of three disc galaxies (IC~2531, NGC~4013 and UGC~2048).  The
  subtraction from the images of a model representing the axisymmetric
  component of the galaxies reveals the contribution of the non-axisymmetric
  terms. We find a direct correspondence between the orbital profiles of 3D
  bars in models and the observed main morphological features of the
  residuals. We also apply a simple unsharp masking technique in order to
  study the sharpest features of the images. Our basic conclusion is that the
  morphology of the boxy ``bulges'' of these galaxies can be explained by
  considering disc material trapped around stable 3D periodic orbits. In most
  models these building-blocks periodic orbits are bifurcated from the planar
  central family of a non-axisymmetric component, usually a bar, at {\em low
    order} vertical resonances. In such a case the boxy ``bulges'' are {\em
    parts} of bars seen edge-on. For the three galaxies we study the families
  associated with the ``peanut'' or ``X''-shape morphology are most probably
  bifurcations at the vertical 2/1 or 4/1 resonance.
\end{abstract}

\begin{keywords}  
  galaxies: kinematics and dynamics; ISM: dust, extinction -- Infrared:
  galaxies -- Infrared: ISM 
\end{keywords} 

\section{Introduction}\label{intro}
In Patsis \& Grosb{\o}l (1996), Patsis, Athanassoula, Grosb{\o}l et al.
(2002a), and Patsis, Skokos \& Athanassoula (2002b, hereafter PSA), have
been presented boxy orbital profiles of 3D, time-independent, models of disc
galaxies. They have been built by combinations of stable, 3D, orbits belonging
in most cases to families of periodic orbits bifurcated from the planar x1
family (see e.g.  Contopoulos 2002, or Contopoulos \& Grosb{\o}l 1989) at the
vertical $n/1$ resonances.  The most efficient dynamical mechanism for
introducing vertical resonances in a system is the presence of
non-axisymmetric components, especially bars.

Vertical profiles of 3D bars have been constructed by PSA, based on the
orbital analysis of 3D Ferrers bars (Pfenniger 1984, 1985; Skokos, Patsis \&
Athanassoula 2002a,b). A profile of a single family includes stable orbits in
a range of ``energies'' (Jacobi constants - $E_j$). In order to construct a
family profile one needs to know its stable orbits. Out of this library of
stable periodic orbits any subgroup that helps in matching the observed
morphological structure can be chosen. Every family has its own characteristic
orbital profile, and the profiles of the same family in several models are
similar. Combinations of the profiles of several families of a model comprise
a model profile. Particular useful representations of the orbital profiles
that allow comparisons with real galaxies, or with snapshots of $N$-body
simulations, are their {\em weighted images} (PSA). The advantage of these
images is that they show the relative importance of every orbit in a profile.
They also show the details of the morphology in locations where more than one
orbit contribute. The images are composed by periodic orbits weighted by the
mean density of the model at the points visited by the orbit. Along each orbit
one picks points at equal time steps. The density of the model is calculated at
each of these points and the mean density is taken as the weight of the orbit.
Having constructed images for every orbit (normalized over its total
intensity), we can combine them to built a profile for a family of orbits. For
the profile of a family are used orbits equally spaced in their mean radius.

The basic conclusions of the orbital analysis relevant to the present study
are:
\begin{enumerate}
\item The morphology supported by the composite orbital profile of a family
  may differ from the morphology of the individual periodic orbits calculated
  at a particular $E_j$. Thus, in
  order to get the backbone of a structure to be compared with the morphology
  of a galaxy, one needs to know the evolution of the shape
  of the stable orbits of a family as a function of 
  $E_j$. 
  
\item The radial extent of the profile of an orbital 3D family is usually
  confined within a radius corresponding to a certain $E_j$ value. Orbits with
  a 
  Jacobi constant larger than this $E_j$, increase their size by increasing
  practically only in the vertical dimension, and this leads to models with
  stair-type edge-on profiles (for more details see Patsis et al. 2002a).
\end{enumerate}

To these two, one should add that it is the presence of the vertical
resonances and not the detailed kind of perturbation in the models that shapes
the boxiness of the orbital profiles. However, ``X''-like features as these we
discuss here are typical of strong bar components.

Recently, unsharp masking techniques applied to images of galaxies (Aronica,
Athanassoula, Bureau et al. 2003; Aronica, Bureau, Athanassoula et al. 2004;
Bureau, Athanassoula, Chung et al. 2004) as well as to images of snapshots of
$N$-body simulations (Athanassoula 2005a,b) have shown excellent agreement
between the image morphologies and what is predicted by the orbital theory in
PSA.  In particular, Aronica et al. (2003), compared the image of ESO~597-036
after unsharp masking with a model in PSA. They found besides a conspicuous
``X''-shape feature in the central part, surface brightness enhancements along
the equatorial plane of the galaxy. Both features have their counterparts in
the orbital models and can be explained by material trapped around stable
periodic orbits. Similar features are indicated by Aronica et al. (2004) also
for ESO-443G042 and by Bureau et al. (2004) for the case of NGC~128. The
comparison of features between orbital and $N$-body models on the other hand
show agreement in even finer details. In both papers by Athanassoula (2005a,b)
one can find in the edge-on views of the models besides ``X''-like features,
density enhancements on the equatorial plane and features like ``parentheses''
(see e.g. Fig.~6 in Athanassoula 2005a). This is a strong indication that a
large percentage of the material in the $N$-body simulation follows orbits
around the families bifurcated at the vertical $n/1$ resonances with small
$n$.

In the present paper we apply two image processing techniques on the images of
three edge-on disc galaxies with rather boxy ``bulges'' in order to detect
structures similar to those predicted by the orbital theory. We focus our
attention to the structures observed in the central regions of IC~2531,
NGC~4013 and UGC~2048 (NGC~973). First we subtract from the I-band images of the
galaxies an axisymmetric model that was found by Xilouris, Kylafis,
Papamastorakis et al. (1997) to describe best the smooth distribution of
stars and dust in these objects. The models are used to isolate the
non-axisymmetric term in the profiles of the galaxies. This term is expected
to reflect in a straightforward way the presence of vertical resonances (PSA).
A second technique we apply is just a gaussian filtering (unsharp masking) on
the images. In Section \ref{observations} we give information about the
observations of the galaxies, in Section \ref{model} we describe the different
image processing techniques, in Section \ref{results} we give the results of
the image processing we performed and finally we discuss our results and
present our conclusions in Section \ref{conclusions}.

\section{Observations}\label{observations}
The observations and the data reduction of the galaxies analyzed in this study
are presented in Xilouris, Kylafis, Papamastorakis et al. (1997), and
Xilouris, Byun, Kylafis et al. (1999). We briefly repeat here both.

Observations of the galaxies UGC 2048 and NGC 4013, were made at Skinakas
observatory in Crete, using the 1.3 m telescope, where a Thomson $1024 \times
1024$ CCD camera is installed at the prime focus of the $f/7.7$
Ritchey-Cretien telescope. The 19$\mu$m pixels of this camera correspond to
0.39 arcseconds on the sky giving a total field of view $6.7' \times 6.7'$.
IC 2531 was observed with the 1 m Australian National University telescope
(ANU) at the Siding Spring Observatory. The CCD camera in this case is an EEV
$576 \times 380$ giving a pixel size of 0.56 arcseconds at the $f/8$
Cassegrain focus. The I passband comparable to that of the Cousin's
photometric system is used in both cases. The exposure times are 20, 25 and 25
minutes for UGC~2048, NGC~4013 and IC~2531 respectively.

In Table 1 we summarize the basic
observational properties of the galaxies as well as some characteristic
parameters describing the axisymmetric stellar and dust components (see
Section \ref{model}). 
 
Out of the samples of the galaxies in Xilouris et al. (1997) and  Xilouris et
al. (1999), for which models for the axisymmetric light and dust distribution
exist, we have selected three of them with boxy profiles to apply our image
processing techniques.
%ttttttttttttttttttttttttttttttttttttttttttttttttttttttttttttttttttttttt
\begin{table*}
\caption{Observational data for the galaxies together with a set of parameters
  describing the axisymmetric stellar and dust components
(see the text in Section \ref{model} for a detailed description of these
parameters).} 
\begin{tabular}{lcccccccccccc}
\hline
Name & Pixel size  & Exp. time & $D$ & Inclination & $h_s$ & $h_d$ & $z_s$
& $z_d$ & $R_e$ & $I_s$ & $I_b$ \\
     & \arcsec    &  (min)  & (Mpc)& (degrees) & (kpc) & (kpc) & (kpc) &
     (kpc) & (kpc) & (mag$/\sq^{\arcsec}$) & (mag$/\sq^{\arcsec}$) & \\
\hline
UGC 2048 & 0.39  & 20 & 63 & 89.6 & 11.0 & 16.5 & 1.0 & 0.57 & 2.4 &18.4 & 9.1 \\
NGC 4013 & 0.39  &  25 & 12 & 89.6 & 1.8 & 2.6 & 0.2 & 0.10 & 1.6 & 17.3 & 10.5 \\
IC 2531 & 0.56  &  25 & 22 & 89.7 & 5.0 & 8.4 & 0.4 & 0.20 & 1.6 & 18.2 & 11.0 \\
\hline
\end{tabular}
\end{table*}
%ttttttttttttttttttttttttttttttttttttttttttttttttttttttttttttttttttttttt
\section{Determining the underlying structure of the galaxy}\label{model} 
Disc galaxies are not axisymmetric systems. Their surface brightness can be
described as the sum of an axisymmetric and a perturbing term. The latter
refers to the presence of spirals or bars. The axisymmetric component is the
most important and can be
described by simple smooth functions (like axisymmetric exponential discs for
the stars and the dust in the plane of the galaxy and a de Vaucouleurs
$R^{1/4}$ law for the bulge). Besides the global perturbations we have clumps
and star forming regions that cause deviations from smoothness locally.  In
the present paper we are interested in the global non-axisymmetric
components.

%Usually the light of a barred disc galaxy can be decomposed in a disc, a
%bulge and a bar component. Particularly instructive is the analysis of images of
%edge-on disc galaxies by L\"{u}tticke, Dettmar and Pohlen (2000). There is a
%lot of ongoing discussion about whether boxy bulges are distinct components or
%pseudobulges, i.e. central features built by secular processes (Kormendy \&
%Kennicut 2004).

Studying the smooth distributions is not only instructive (one gets a very
good description of the galaxy with as few parameters as possible) but can
also be a very helpful tool to uncover features due to non-axisymmetric
components (e.g. bars). This is done by subtracting the axisymmetric model
from the image of the galaxy. If the axisymmetric model is able to describe
well the large-scale structure, then, with this method, what it left, i.e. the
residuals, will indicate features that mainly have to do with either the
non-smooth distribution of the dust (seen as absorption features) or the
non-axisymmetric component of the stars (seen as excess of starlight). Assuming
a constant mass--to--light ratio (M/L), this light excess will indicate
regions of enhanced stellar density and will be comparable with the orbital
models. 
Good agreement will indicate that large percentages of stars are trapped
around stable periodic orbits.

The model that we use is described in great detail in Xilouris et al. (1997).
The stellar emissivity (luminosity per unit volume) that we use consists of an
exponential (in both radial and vertical directions) disc and a bulge
described by the $R^{1/4}$ law, namely
\begin{displaymath}
L(R,z) = L_s \exp \left( - \frac{R}{h_s} - \frac{|z|}{z_s} \right)
\end{displaymath}
\begin{equation}
~~~~~~+ L_b \exp (-7.67 B^{1/4}) B^{-7/8},
\end{equation}
with $h_s$ and $z_s$ being the scalelength and scaleheight of the
disc and
\begin{equation}
B = \frac{\sqrt{R^2 + z^2 (a/b)^2}}{R_e} ,
\end{equation}
with $R_e$ being the effective radius of the bulge and $a$ and $b$ the
semi-major and semi-minor axis respectively. Here $L_s$ and $L_b$ are the
normalization constants for the stellar emissivity of the disc ($L_s$) and the
bulge ($L_b$).
The central value for the surface brightness of the disc and
the bulge, if the model galaxy is seen edge-on and there is no dust, are given
by $I_s = 2L_sh_s$ and $I_b = 5.12L_bR_e$.

For the extinction coefficient we use a double exponential law, namely
\begin{equation}
\kappa_{\lambda}(R,z) = \kappa_{\lambda}
\exp \left(- \frac{R}{h_d} - \frac{|z|}{z_d}
\right) ,
\end{equation}
where $\kappa_{\lambda}$ is the extinction coefficient at
wavelength $\lambda$ at the center of the disc and
$h_d$ and $z_d$ are the scalelength and scaleheight respectively
of the dust. 

The radiative transfer model that we have used is the one described by Kylafis
\& Bahcall (1987) (see also Xilouris et al. 1997). The results are summarized
in Table 1, where we give the values of the most important parameters in the
best fittings. 

The unsharp masking has been done by means of standard ESO-MIDAS commands.
First we apply on the image a gaussian filter with radius of 5 pixels around
the central pixel and with mean and sigma values of 5 and 2 pixels
respectively in both directions. With this procedure a blurred image of the
galaxy is created. Then, after subtracting the filtered from the original
image the residual indicates the presence of the sharpest features on the
images.

\section{Results}\label{results}
In Fig.~1 we show the residual maps created by subtracting the model images of
Xilouris et al. (1997, 1999) from the observed images of the three galaxies
(NGC~4013 (a), IC~2531 (b) and UGC~2048 (c)). The images have been rotated so
that the major axis of the galaxies is along the large dimension of the frame.
The difference between the models and the observations are relatively small,
indicating the ``perturbation'' character of the non-axisymmetric components
(on average 15\% - Xilouris et al. 1997, 1999). However, this procedure offers
a straightforward tracer exactly of this excess light we are interested in.
%ffffffffffffffffffffffffffffffffffffffffffffffffffffffffffffffffffffff
\begin{figure*}
\centerline{\includegraphics[scale=1]{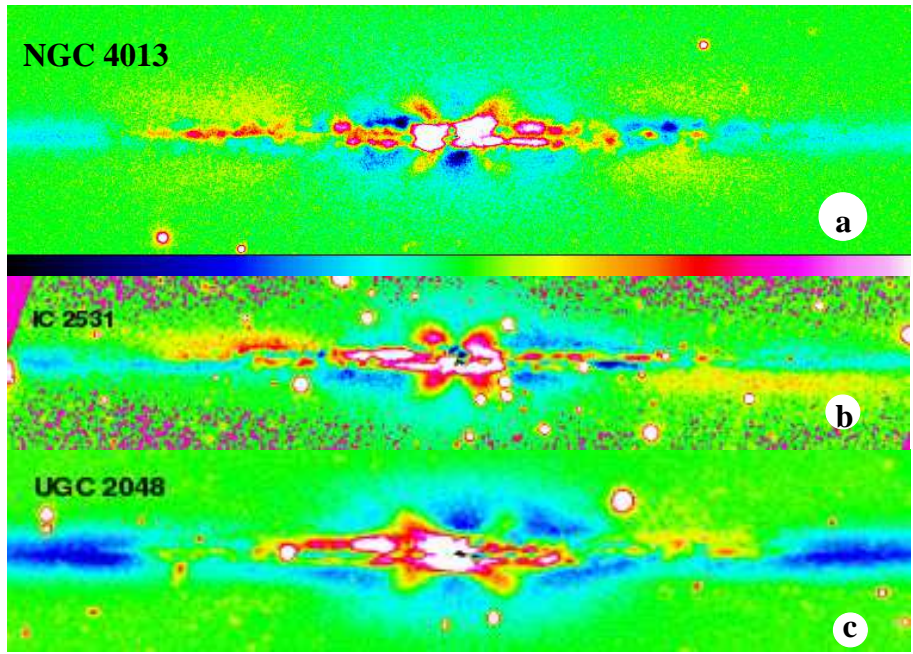}}
%\vskip.2in
\caption{Residual images for NGC~4013 (a), IC~2531 (b) and UGC~2048 (c). From
  the original images it is subtracted a best fitting axisymmetric profile
  according to the models of Xilouris et al. (1997). An ``X''-shaped feature
  embedded in the boxy ``bulge'' is conspicuous in all three cases. Color
  coding can be seen below (a).}
\end{figure*}
%fffffffffffffffffffffffffffffffffffffffffffffffffffffffffffffffffffffff
The main characteristic feature is an ``X''-shaped structure revealed in the
central parts of the galaxies. By comparing the residual with the full images
of the galaxies we observe that the ``X'' features are in the region, where we
have the boxy bulges.

The images after unsharp masking show sharp features that resemble the wings
of an ``X''. This can be seen in Fig.~2 for the same galaxies.
%fffffffffffffffffffffffffffffffffffffffffffffffffffffffffffffffffffffff
\begin{figure*}
\centerline{\includegraphics[scale=0.6]{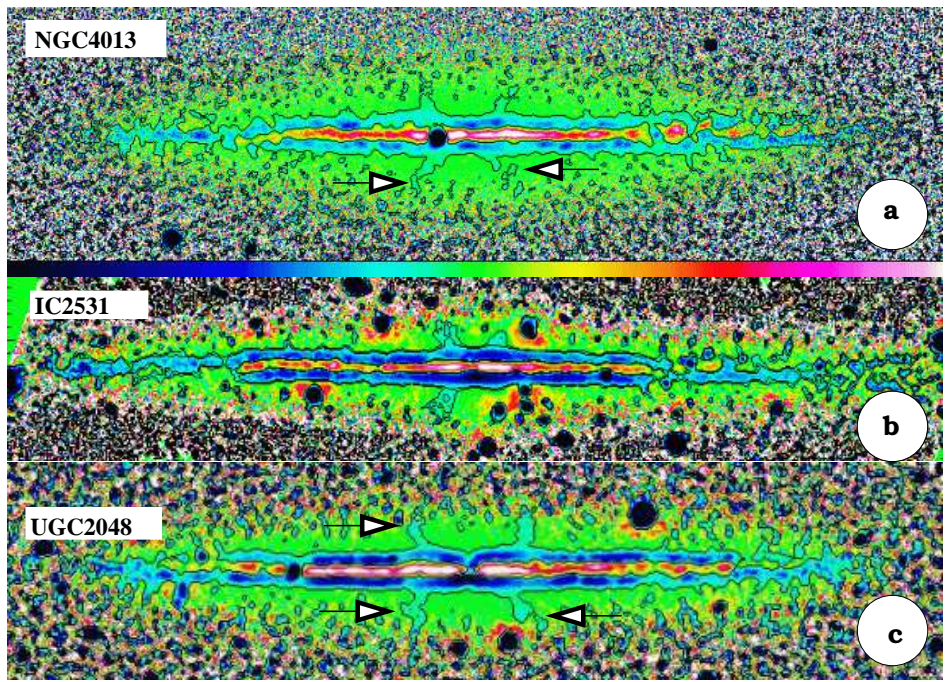}}
%\vskip.2in
\caption{The result of unsharp masking on the images of NGC~4013 (a), IC~2531
  (b) and UGC~2048 (c). ``X'' shape features similar to the structures 
  foreseen by the orbital models can be found in all three cases. Arrows
  indicate breaks of the ``X'' wings. Color
  coding can be seen below (a).}
\end{figure*}
%fffffffffffffffffffffffffffffffffffffffffffffffffffffffffffffffffffffff
Characteristic contours underline the main structure observed. In this case we
can better distinguish the highest contrast features. Again they are located
at the region occupied by the boxy bulges. They should be compared to features
expected in the central parts of 3D bars viewed almost, edge-on. The
``X-shape'' is encountered only in models for bars, while boxy bulges in
general can be found also in non-barred models (Patsis et al. 2002a).

\section{Discussion and Conclusions}\label{conclusions}
Both image processing techniques revealed a kind of ``X'' structure in the central
components of the three edge-on galaxies with boxy profiles we studied.  In
Fig.~1, where we give the residuals after subtracting a Xilouris et al. (1997)
model we see the overall shape of the non-axisymmetric component. The unsharp
masking 
in Fig.~2 shows more clearly the high intensity features of the ``bulges''.
They emerge out of the equatorial plane as distinct branches.

In most cases of the models in PSA an ``X'' is supported by material trapped
around x1v1 orbits (Skokos et al. 2002a), i.e. by stable 3D orbits introduced
in the system at the vertical 2/1 resonance. However, there are other 3D
families as well, which could support straight line segments emerging out of
the equatorial plane in the edge-on projections of the orbital models. Such is
the case of x1v5, according to the nomenclature of Skokos et al. (2002a),
which is a family born at the vertical 4/1 resonance. This family has been
associated with boxy central components already by Pfenniger (1985). A typical
orbital profile is given in Fig.~7a in PSA. 
%\newline

In models (orbital or $N$-body), a visual difference can be found if we
compare the side-on views. The distance between the wings of the ``X'' feature
in the x1v5 case is larger than the distance of the corresponding wings in
x1v1 profiles. The difference can be seen in Fig.~3, where we give in (a) a
side-on typical x1v1 profile and in (b) a profile dominated by the presence of
the x1v5 family. They correspond to models
D and B in PSA.  As we mentioned in the introduction, there is a certain value
of $E_J$, beyond which the orbits of a family grow practically only in the $z$
direction. So, for each family, there is a maximum radius on the equatorial
plane, within which the projections of its orbits are confined.  In the
side-on views we can measure the length of the projection along the major axis
of the bar.  In the two specific examples we give here, the projection of the
x1v1 orbits on the major axis of the bar in Fig.~3a corresponds roughly to a
length 45\% of the longest bar supporting orbits, i.e. the length of the bar
in our orbital model.  On the other hand in Fig.~3b, this percentage
reaches almost 90\%.
%fffffffffffffffffffffffffffffffffffffffffffffffffffffffffffffffffffffff
\begin{figure}
\centerline{\includegraphics[scale=0.58]{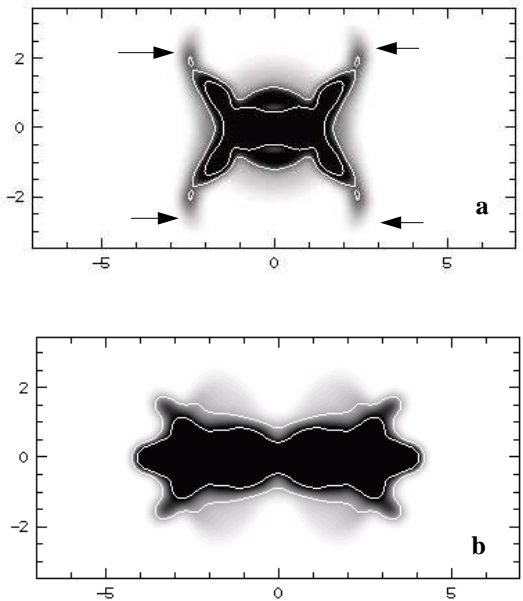}}
\caption{(a) A typical x1v1 side-on profile based on orbits of model D in
  PSA. Arrows indicate the inward bending of the ``X'' wings
  at large distances above the equatorial plane. (b) A typical side-on x1v5
  profile corresponding to model B in PSA. It extends to
  larger radii than the profile in (a).}
\end{figure}
%fffffffffffffffffffffffffffffffffffffffffffffffffffffffffffffffffffffff
Unfortunately this criterion cannot in general apply on the images of real
galaxies since we are missing the essential information about the orientation
of the boxy structure with respect to the line-of-sight. Thus, at the level of
the current analysis, we cannot point to a single vertical resonance and
attribute to it the observed morphological feature. 
%In the cases we
%study, the distance between the wings of the ``X'' feature are larger on the
%image of UGC~2048. 
A galaxy may lack a vertical 2/1 (or even 3/1) resonance,
in which case material may be trapped by stable orbits bifurcated at the
vertical 4/1 resonance. The corresponding profile in such a case will look
like what we see in Fig.~3b. It consists of stable orbits of the following
families: (a) x1v5, (b) a bifurcation of it, and (c) z3.1s (PSA). 
%It is
%remarkable that this model has a clear minimum at the center of the galaxy, as
%the isocontours of the UGC~2048 image after unsharp masking (Fig.~2c).

Another point that has to be mentioned, is the inwards bending or break of
the wings of ``X'' close to their maximum height above the equatorial plane.
According to the orbital models this can be explained by the fact that the
orbits of a family increase their size practically only in the vertical
direction beyond a certain $E_j$ value. In Fig.~3a we point with arrows to
these breaks at the x1v1 profile. However, this is expected to happen to the
profiles of other families as well if they are populated with orbits with high
energies ($E_j$) (see a characteristic case in PSA, Fig.18). This tendency can
be observed at the wings of the ``X'' features in Fig.~2a (NGC~4013) and
Fig.~2c (UGC~2048). We indicate these breaks with arrows.

The main conclusion of the present study is that the careful subtraction of
axisymmetric components from the profiles of the three edge-on disk galaxies
(NGC~4013, IC~2531 and UGC~2048) reveals an ``X''-like morphology, which
indicates the presence of a bar. With this method we could isolate the light
coming from the non-axisymmetric components. The similarity of these components
with the morphology of the orbital profiles is a strong indication that the
boxy ``bulges'' of these three galaxies are parts of bar structures observed
edge-on.  If the ``X''-like structure of the residuals is due to stars trapped
around stable periodic orbits, then we expect the presence of sharp features
(the wings of the ``X'') {\em inside} the bar region. By applying the unsharp
masking on the images we find, in the foreseen by the orbital models regions,
sharp features. The two methods are complementary. The unsharp masking does
not give any information about the overall shape of the non-axisymmetric
component. It shows however that in the regions we see the residuals we have
the expected sharp features. These features show a nice correspondence with
the dense parts of published orbital models (see PSA, Fig.1a, 3b, 7a, 9a).
These techniques, and especially the study of the residuals after subtracting
axisymmetric models, should be applied in a large sample of galaxies in order
to estimate the frequency of these features in the profiles of boxy edge-on
galaxies.

\section{Acknowledgments}
This work was partly supported by the Research Committee of the Academy of
Athens. 
It is a pleasure to acknowledge fruitful discussions and valuable
comments by G.~Contopoulos and N.~Kylafis. We thank the referee, Michael
Pohlen, for very useful comments that helped us improve the paper.

\end{document}